\documentclass[aps,pre,twocolumn]{revtex4}
\usepackage{amssymb}

\usepackage{eurosym}
\usepackage{epsfig}

\begin{document}

\title{Anisotropic characteristics of the Kraichnan direct cascade in
two-dimensional hydrodynamic turbulence}
\author{E.A. Kuznetsov$^{a,b,c}$\/\thanks{%
kuznetso@itp.ac.ru} and E.V. Sereshchenko$^{a,d,e}$}
\affiliation{{\small \textit{$^{a}$ Novosibirsk State University, 630090 Novosibirsk,
Russia} }\\
{\small \textit{$^{b}$ Lebedev Physical Institute, RAS, 119991 Moscow, Russia%
}}\\
{\small \textit{$^{c}$ Landau Institute for Theoretical Physics, RAS, 119334
Moscow, Russia}}\\
{\small \textit{\ $^{d}$ Khristianovich Institute of Theoretical and Applied
Mechanics, SB RAS, 630090 Novosibirsk, Russia}}\\
{\small \textit{$^{e}$ Far-Eastern Federal University, 690091 Vladivostok,
Russia }}}

\begin{abstract}
Statistical characteristics of the Kraichnan direct cascade for two-dimensional hydrodynamic turbulence
are numerically studied (with spatial resolution $8192\times 8192$) in the presence of pumping and viscous-like
damping. It is shown that quasi-shocks of vorticity and their Fourier partnerships in the form of jets introduce
an essential  influence in turbulence leading to strong angular dependencies for correlation functions. The energy
distribution as a function of modulus $k$ for each angle in the inertial interval has the Kraichnan behavior,
$\sim k^{-4}$, and simultaneously a strong dependence on angles. However, angle average provides with a high accuracy
the Kraichnan turbulence spectrum $E_k=C_K\eta^{2/3} k^{-3}$ where $\eta$ is enstrophy flux and the Kraichnan
constant $C_K\simeq 1.3$, in correspondence with the previous simulations. Familiar situation takes place for third-order
velocity structure function $S_3^L$ which, as for the isotropic turbulence, gives the same scaling with respect to
separation length $R$ and $\eta$, $S_3^L=C_3\eta R^3$, but the mean over angles and time  $\bar {C_3}$ differs from its
isotropic value.
\end{abstract}

\maketitle

\vspace{0.5 cm}

\section{Introduction}
Developed two-dimensional hydrodynamic turbulence, in contrast to three-dimensional turbulence,
in the inertial interval of scales possesses an additional integral of motion --- enstrophy that is a half  integral of the squared vorticity, $(1/2)\int_{s}\Omega ^{2}d\mathbf{r}$. As was demonstrated in 1967 by Kraichnan \cite{kraichnan},
the existence of this integral generates in the inertial range its own Kolmogorov-type spectrum of turbulence,
\begin{equation}
E(k)~\sim ~k^{-3},  \label{Kraichnan}
\end{equation}%
(now called the Kraichnan spectrum). This spectrum corresponds to a constant enstrophy flux toward the small-scale
region. Simultaneously, according to \cite{kraichnan}, a conventional Kolmogorov spectrum $E(k)~\sim ~k^{-5/3}$ is formed at 
large scales with a constant energy flux directed toward the region of small values of $k$ (inverse cascade). Since the 
Kraichnan paper in 1967 there were performed many  numerical experiments (see \cite{lilly} - \cite{okhitani};
a more complete list in \cite{KNNR-07}) testified in favor of existence of the Kraichnan spectrum.
Already in the first numerical experiments (see, e.g., \cite{lilly}) there was observed the emergence of sharp
vorticity gradients (consistent with the high Reynolds number). It corresponds to the formation of the vorticity jumps similar
to shock waves with thickness small compared with their length. Based on these observations, Saffman \cite{saffman}
proposed another spectrum, $E(k)~\sim ~k^{-4}$. The Saffman spectrum was obtained under the assumption that the main
contribution to the spectrum comes from isotropically distributed vorticity shocks. (It should be noted that in
two-dimensional turbulence, the formation of singular vorticity gradients in finite time is forbidden, in accordance
with the rigorous theorems \cite{wolibner}; therefore we will call such structures as quasi-shocks.) On the other hand,
it follows from simple considerations that the spectrum with such shocks is expected rather to have the Kraichnan type
behavior than that predicted by Saffman. The Fourier amplitude $\Omega _{k}$ from one shock will be $\propto k^{-1}$,
that immediately yields the spectrum $E(k)~\sim ~k^{-3}$. However, the situation is not too simple. If one assumes
that the characteristic length of the step $L\gg k^{-1}$, then the energy distribution from one such step in the
$k$-space has the form of a jet with an apex angle of the order of $\left( kL\right) ^{-1}$ . For freely decay
turbulence as was shown numerically in \cite{KNNR-07,KNNR-10,KKS-13} the spectrum (\ref{Kraichnan}) arises owing
to vorticity quasi-shocks, which form a system of jets with weak and strong overlapping in the $k$-space. In their
turn, a reason of appearance of quasi-shocks can be regarded due to the compressible features of the so-called
di-vorticity, $\mathbf{B=\nabla \times }\Omega \widehat{z}$. For inertial scales this vector field is frozen-in-fluid
and its value, in the general case, according to \cite{KNNR-07,KNNR-10,KKS-13,K-04} can be changed due to the velocity
component $\mathbf{v_{n}}$, normal to the $\mathbf{B}$-line, for which $\mbox{div}\,\mathbf{v_{n}}\neq 0$. This is the
origin of compressibility for continuously distributed di-vorticity lines and thereby appearance of quasi-shocks in 2D.
Besides, for freely decay turbulence numerically it was established that (i) in each of such jets, the fall in $E(k)$
at large $k$ is proportional to $~k^{-3}$, which after the angle averaging yields the spectrum (\ref{Kraichnan}) ;
{(ii) the third-order velocity structure function $S_{3}^{L}(\mathbf{R})=\langle \delta v_{\Vert}^{3}\rangle $ depends
proportionally to $R^{3}$ (where $\mathbf{R}=\mathbf{r}-\mathbf{r}^{\prime }$, \ $\delta v_{\Vert }$ the projection of
the velocity difference onto the vector $\mathbf{R}$), in complete agreement with the Kraichnan theory.}

The main aim of this paper is to study numerically in which extent all the above statistical properties observed in
freely decay two-dimensional turbulence remain for the direct (Kraichnan) cascade when the presence of both pumping
and viscous damping. Here we use the same numerical scheme reported in our previous paper \cite{KKS-13} with
$8192\times 8192$ grid points and periodical boundary conditions. In order to eliminate an influence of the inverse
cascade we introduce at very large scales (comparable with the box size) a big damping that allows us to increase the
inertial interval for the direct cascade. In the inertial range we put the damping rate equal to zero and switch on
the viscous-like dissipation starting from $k=0.6k_{max}$.

\vspace{0.3cm}
The main results of this paper are as follows:

1. The energy spectrum averaged over angles, as a function of modulars $k$, with a high accuracy has
the Kraichnan behavior: $E(k) = C_K\eta^{2/3}k^{-3}$ where $\eta$ is the enstrophy flux and $C_K$ the Kraichnan constant.
In our simulations $C_K\simeq 1.3$, in correspondence with the previous simulations \cite{Gotoh}-\cite{LindborgVallgren}.

 2. The formation of $k^{-3}$-dependence happens for all rays in the $k$-space starting from
sufficiently earlier evolution times (about two-three characteristic inverse maximal vorticity $\omega_{max}^{-1}$).
However, the angular dependence of energy distribution $\epsilon(\mathbf{k})$, in all our numerical experiments,
remains far from isotropic. In $k_x$-$k_y$ plane it has a very pronounced jet behavior, typical for freely decay
turbulence \cite{KNNR-07, KNNR-10, KKS-13}. At small evolution times when flow is still not-turbulent quasi-shocks
and their respective jets are very visible. With time jets have a tendency to mutual overlapping but
significant angular fluctuations retain in the energy distribution at the steady state.

 3. The jet structures reflect also in the behavior of third-order velocity structure function
$S_3^{(L)}(\mathbf{R})=\langle (\delta v_{\|})^3\rangle$. Like for the isotropic direct cascade with constant enstrophy
flux \cite{Frisch}-\cite{Yakhot} fitting shows that this correlation function behaves linearly on both $\eta$ and $R^3$, $S_3^{(L)}(\mathbf{R})=C_3\eta R^3$, but demonstrates a significant dependence for coefficient $C_3$ on angles $\theta$,
in correspondence with those observed for $\epsilon(\mathbf{k})$. Average of $C_3$ over angles  yields for mean value $\bar{C_3}$
being far from the theoretical prediction for the isotropic turbulence (see, e.g., review \cite{boffetta} and
references therein).

\section{Physical model and numerical scheme}
We consider the two-dimensional Navier-Stokes equation for an incompressible flow $\mathbf{u(x},t)=(u_{x}(x,y,t),u_{y}(x,y,t))$
in the vorticity formulation, $\Omega =\nabla \times \mathbf{u}$,
\begin{equation}
\frac{\partial \Omega }{\partial t}+(\mathbf{u}\nabla )\Omega =F+G\quad
\quad \mbox{\rm with}\quad \quad \emph{div}\text{ }\mathbf{u}=0  \label{NS}
\end{equation}%
supplemented by the periodic boundary conditions in $x$ and $y$ directions in the square box with size $L=1$. Here $F$ is a function
responsible for both injection of the energy and its dissipation on large scales and $G$ for enstrophy
dissipation at large $k$.

To study the direct cascade we model $F$ so that to eliminate influence of the energy condensation due to inverse cascade
by introducing a strong dissipation at $k<k_{p}$ where $k_{p}$ is the characteristic injection wave number. In our
numerics we used function $F$ in the form $F=\hat{\Gamma}\Omega $ where the Fourier transform of  operator $\hat{\Gamma}$
\begin{eqnarray}
\Gamma _{k} &=&A\frac{(k^{2}-b^{2})(k^{2}-a^{2})}{k^{2}}\quad \text{at}\quad
0\leq k\leq b,\quad  \label{Gamma} \\
\quad \Gamma _{k} &=&0\quad \text{at}\quad k>b.
\end{eqnarray}
Parameters $a$ and $b$ were chosen by such a way to get a rapid transition to the stationary energy dissipation at small $k$.

Dissipation function $G=\hat{\gamma}\Omega $ was taken in the viscous-type form:
\begin{eqnarray}
\gamma _{k} &=&0\quad \mbox{\rm at}\quad k\leq k_{c},\quad  \label{gamma} \\
\quad \gamma _{k} &=&-\nu (k-k_{c})^{2}\quad \mbox{\rm at}\quad k>k_{c},
\end{eqnarray}%
where viscous cut-off $k_{c}$ was taken as $0.6k_{\max }$ with $k_{\max}=4096$ being the maximal value of $k$.
Coefficient $0.6$ in front of $k_{\max }$ provide to prevent aliasing.

Appropriate changes of boundaries $k=b$ and $k=k_{c}$ for zeroth values of both $\Gamma _{k}$ and $\gamma _{k}$ allowed us
to get a maximally possible inertial interval for a given spatial resolution. Note that, in the interval $b<k<k_{c}$,
Eq. (\ref{NS}) transforms into the Euler equation. As it was shown analytically and confirmed by numerical experiments
\cite{KNNR-07,KNNR-10, KKS-13,K-04} for freely decay turbulence there exists a tendency for breaking of the di-vorticity
lines. This is consequence of the representation analogous to that for the 3D Euler equation, i.e. the vortex line
representation \cite{KR, kuz}, and thereby the compressible character of the di-vorticity lines. Just this
compressibility is a reason for appearance of sharp vorticity gradients in the form of quasi-shocks and respectively jets
in the $k$-space.

Equation (\ref{NS}) was solved numerically using pseudo-spectral Fourier method, while integration on time was performed
with the help of hybrid Runge-Kutta / Crank-Nicholson third-order scheme. Derivatives was calculated in the spectral
space, whereas the nonlinear terms were determined on a computational grid in the physical space. Convective term was
approximated explicitly, and for linear terms ($F$ and $G$) we used implicit scheme. The transformations from the physical
to the spectral space and back were performed through the Fast Fourier Transform (FFT). Spatial resolution was
$8192\times8192$ points. Simulations were performed (with the use of the NVIDIA CUDA technology) at the Computer Center
of the Novosibirsk State University.

Initial conditions were chosen familiar to those used in our previous paper \cite{KKS-13} as random sets of vortices with
zero mean vorticity. The initial spectrum was concentrated in the region comparable with the injection scale $k_c$.
We verified that variations in the initial conditions qualitatively did not change behavior of the system and its
statistical characteristics.

\section{Numerical results}
First, we determined the parameters for pumping rate $\Gamma $ when turbulence reached its stationary state. These are
$A=0.004$, $a=3$, $b=6$, and $\nu =1.5$. Choice of the first three  parameters provided sufficiently rapid transition
to the stationary state for $k\leq 3$ and combination with the fourth one to approaching stationary asymptotics for the
mean (average over angles) enstrophy flux.

\begin{figure}[b]
\label{Fig1}
\centerline{
\includegraphics[width=0.23\textwidth]{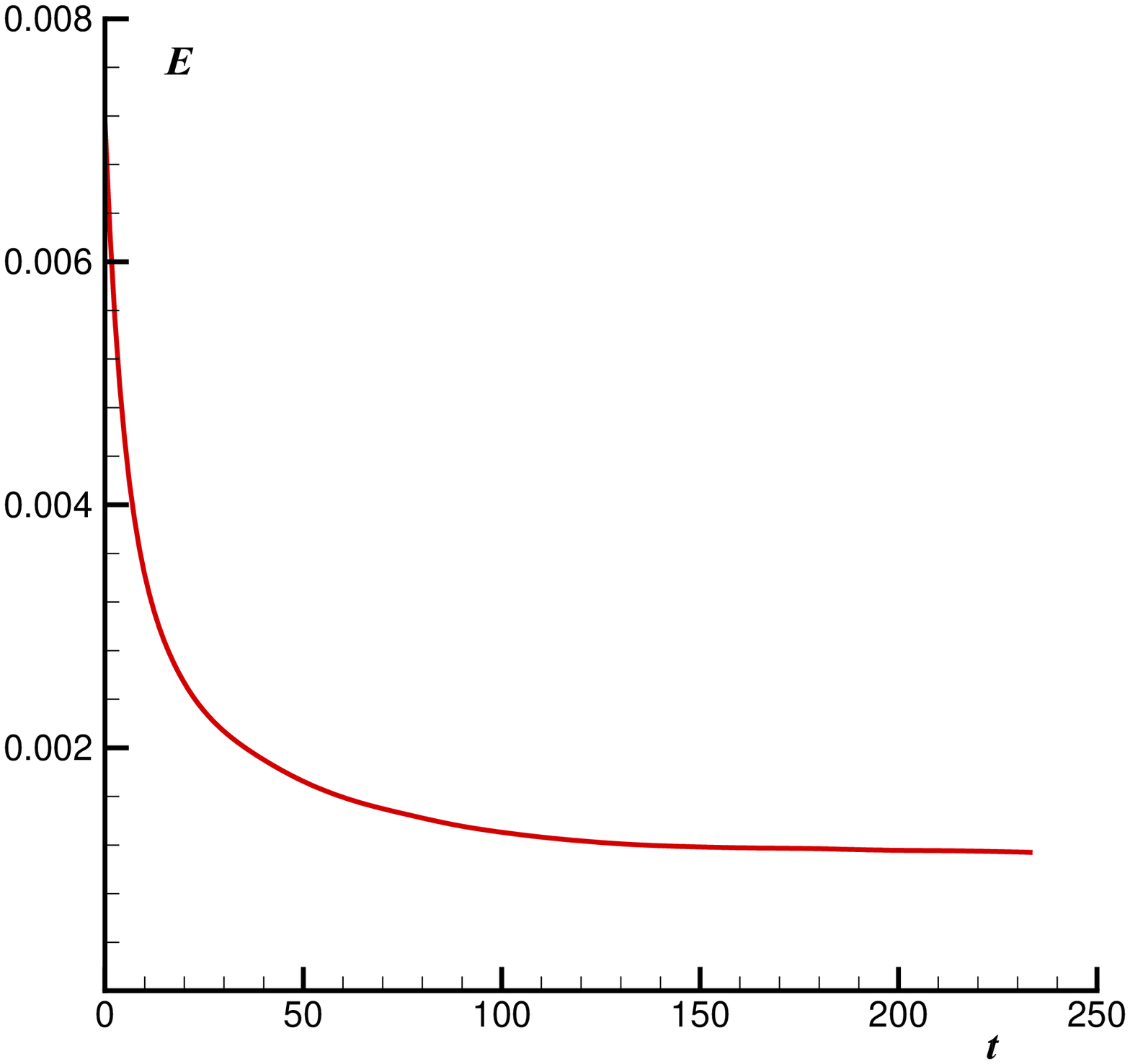}
\includegraphics[width=0.23\textwidth]{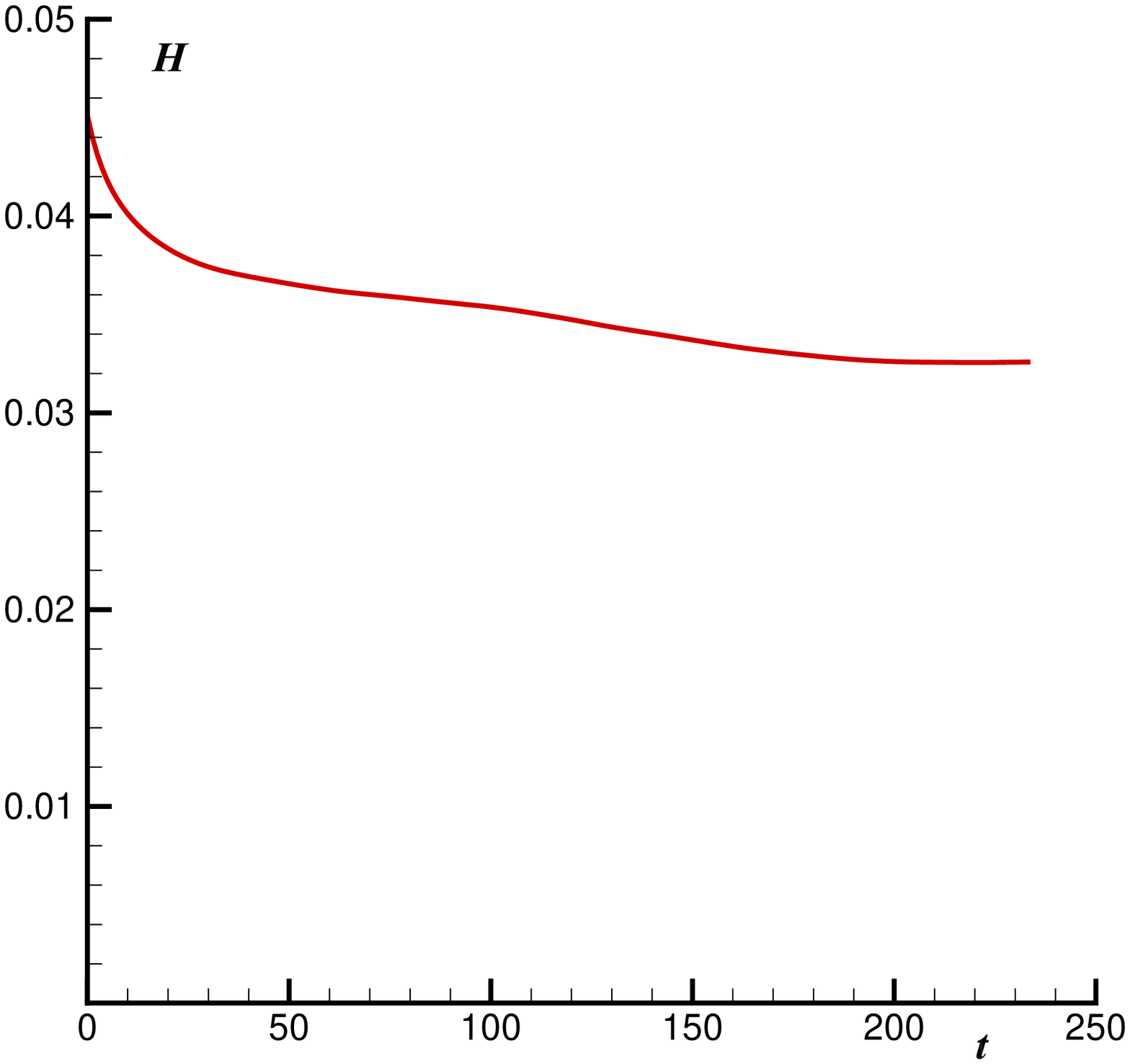}}
\caption{Time evolution of total energy  $E$ and total enstrophy $H$.}
\end{figure}

\begin{figure}[t]
\label{Fig2}
\centerline{
\includegraphics[width=0.45\textwidth]{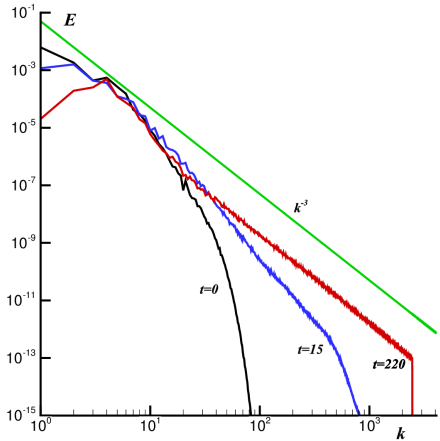}}
\caption{ Energy spectrum $E(k)$ at different instants of time.}
\end{figure}

Fig. 1 for the given parameters $A$, $a$, $b$, and $\nu $ show temporal behavior of the total energy and the enstrophy.
As seen total energy $E$ tends to the constant value (stationary state) at earlier time than the total enstrophy does.
In our numerical experiments we found also how the energy spectrum and both the vorticity and di-vorticity spatial
distributions evolve with time.

Fig. 2 shows the angularly averaged spectrum of turbulence $E(k)$ at different instants of time. At times $t=220$ one can see
 the steady (compare with Fig. 1) spectrum  with the Kraichnan behavior $E(k)\sim k^{-3}$ which is seen almost on
three decades.  At the steady state
($t\gtrsim 200$) by calculating the enstrophy flux $\eta $ as the integral $\int \gamma(k) |\Omega({\bf k}) |^2 d {\bf k}$
we verified that the steady energy spectrum $E(k)$ with a high accuracy coincides with the Kraichnan spectrum
$E(k) = C_K\eta^{2/3}k^{-3}$ where  $C_K\simeq 1.3$ the Kraichnan constant, in correspondence with the previous simulations \cite{Gotoh}-\cite{LindborgVallgren}.

As was shown \cite{KNNR-07,KNNR-10, KKS-13}, for the decay turbulence  such a spectrum is formed owing to the sharp vorticity
gradients appearing due to breaking of di-vorticity lines. Results presented in Figs. 3-5 also show a significant role of
this process on the formation of the Kraichnan spectrum. In Fig. 3 one can see the spatial distributions of $\Omega $ at $t=100$,
before reaching the steady state, and at $t=220$, after its reaching.

\begin{figure}[t]
\label{Fig3}
\centerline{
\includegraphics[width=0.23\textwidth]{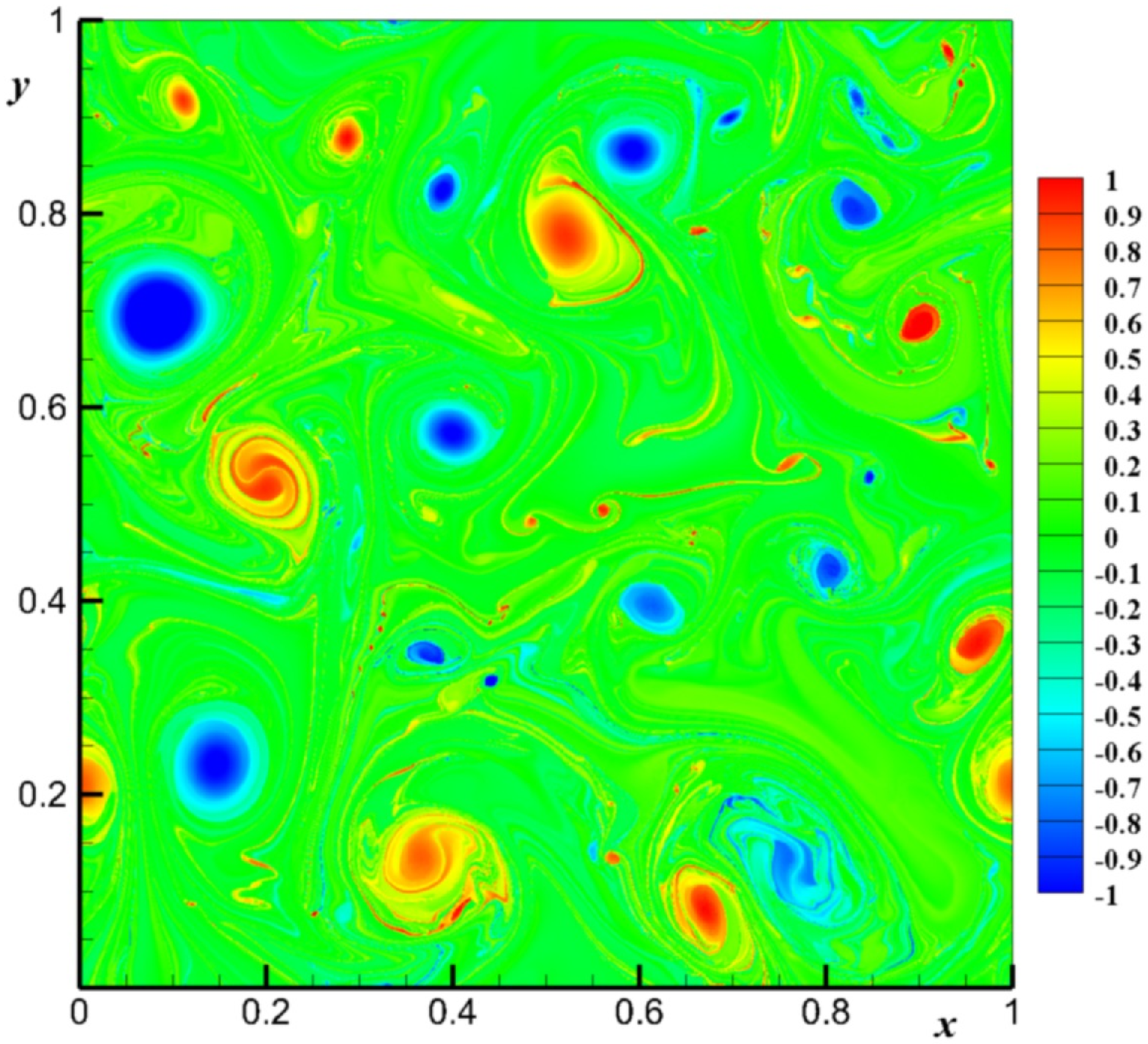}
\includegraphics[width=0.23\textwidth]{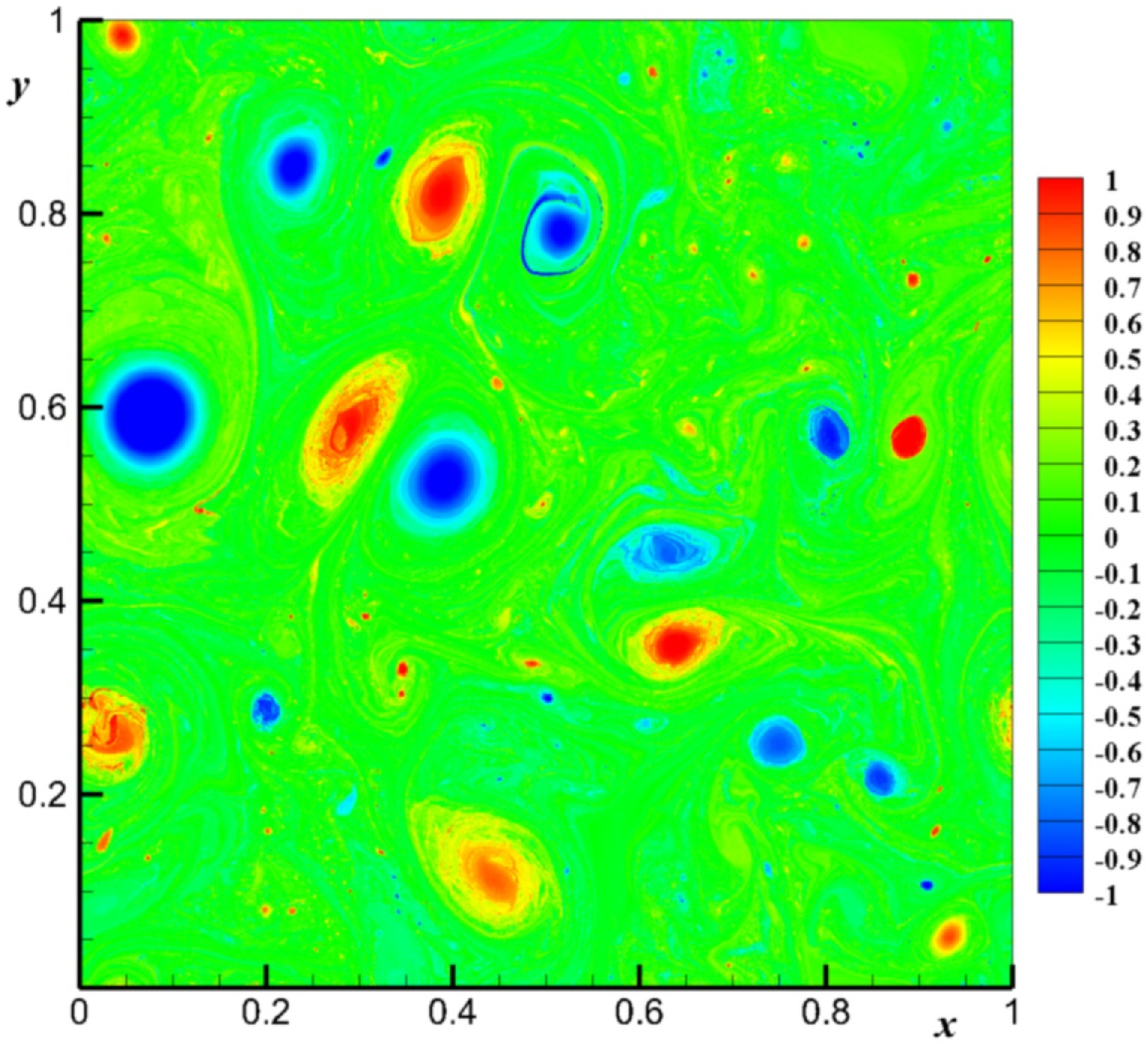}}
\caption{Vorticity distributions at $t=100,220$.}
\end{figure}

\begin{figure}[ht]
\label{Fig4}
\centerline{
\includegraphics[width=0.23\textwidth]{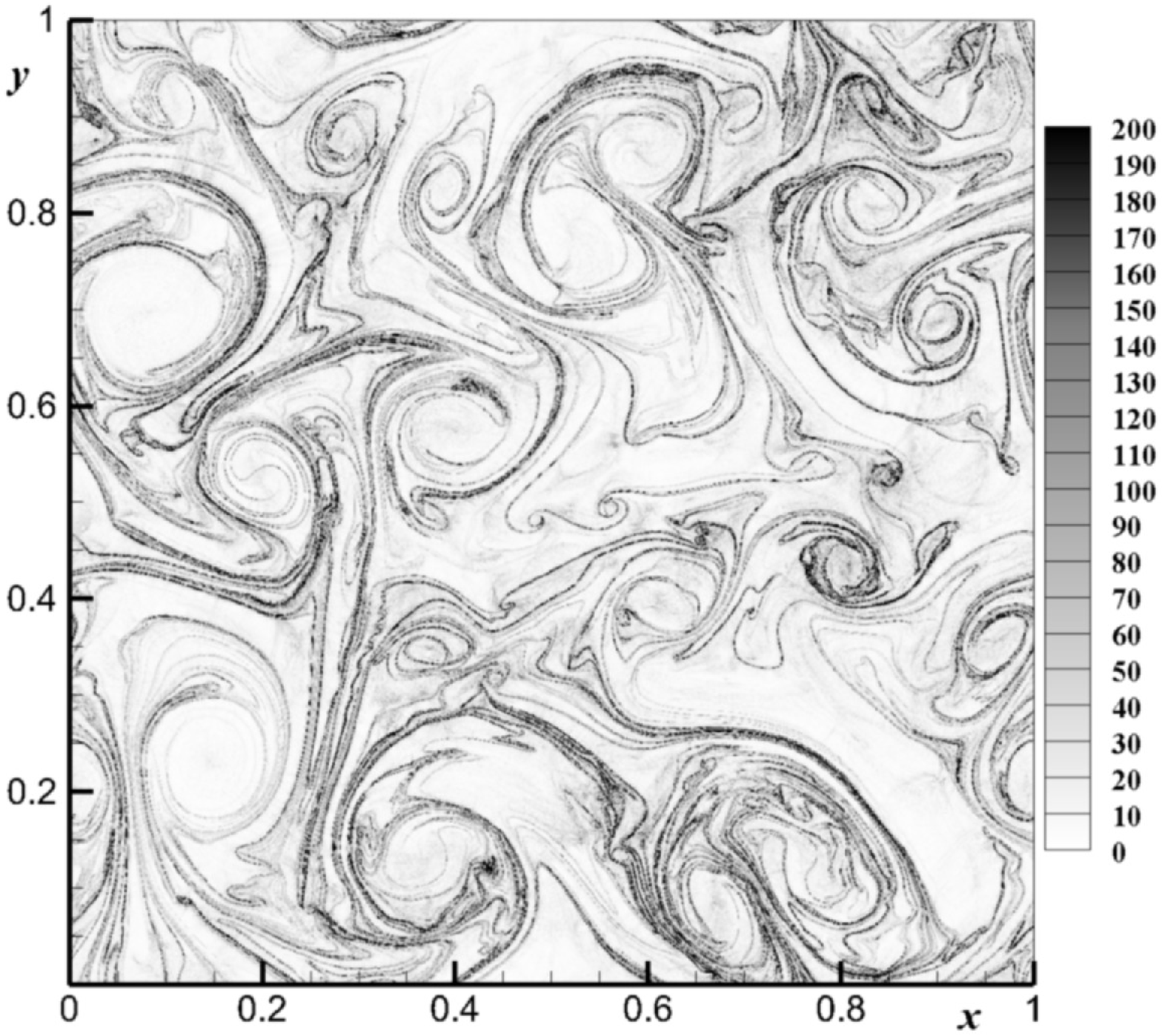}
\includegraphics[width=0.23\textwidth]{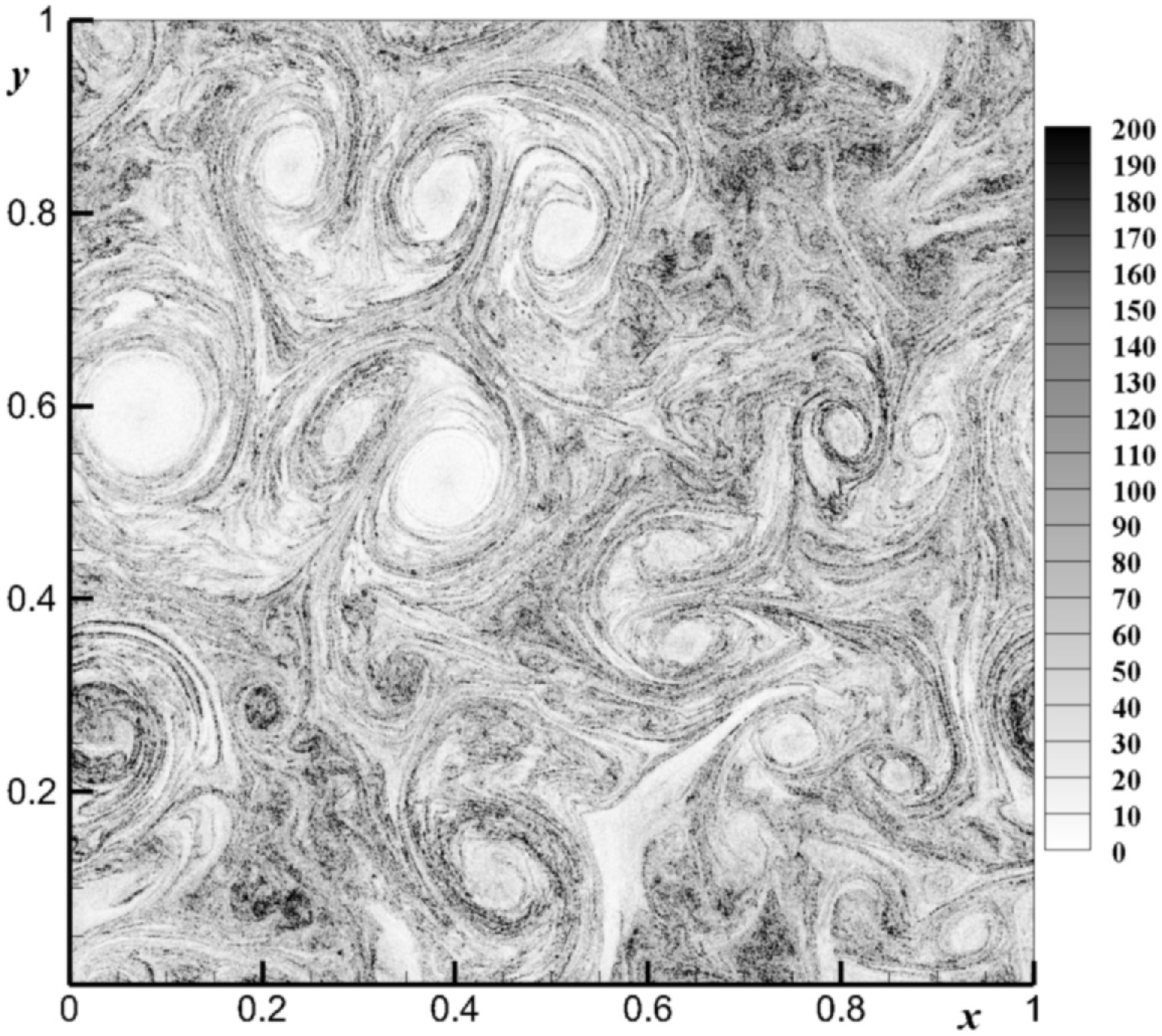}}
\caption{Distribution of $|\mathbf{B}|$ at $t=100,220$.}
\end{figure}

These are typical distributions for $\Omega $, but they are less informative than the corresponding distributions of
di-vorticity $|\mathbf{B(r},t)|$ which are presented in Fig. 4 (for the same time instants as in Fig. 3).

As is seen from these figures the di-vorticity value concentrates along lines which form a whole network, between these lines $|\mathbf{B}|$ is much less. This difference is most clearly seen in the dependence  of $|\mathbf{B}|$ along  projection $y=0.5$
(see Fig. 5)  where maxima of the di-vorticity exceed its minimal values for 1-2 orders of magnitude.

\begin{figure}[b]
\label{Fig5}
\centerline{
\includegraphics[width=0.23\textwidth]{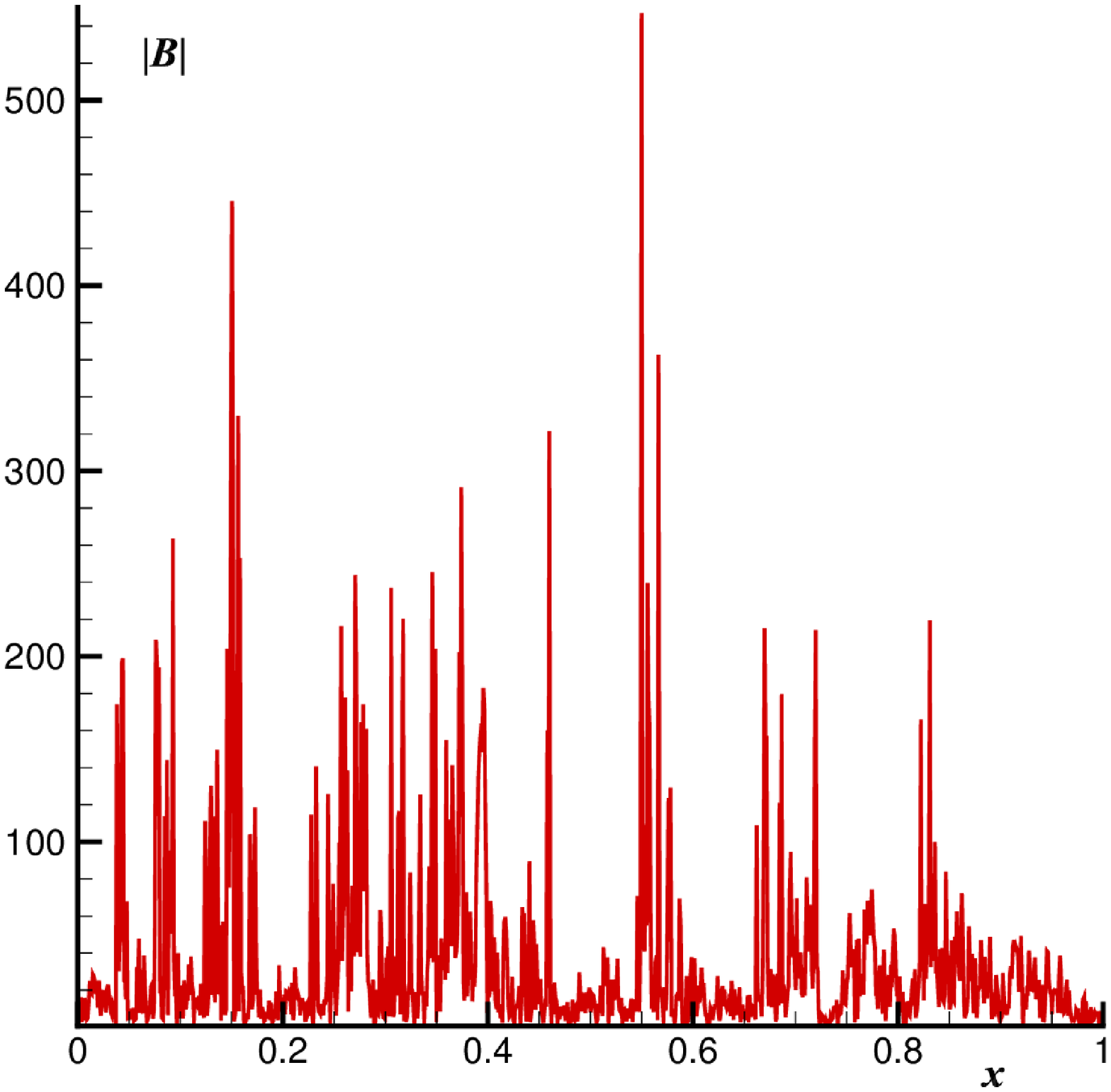}
\includegraphics[width=0.23\textwidth]{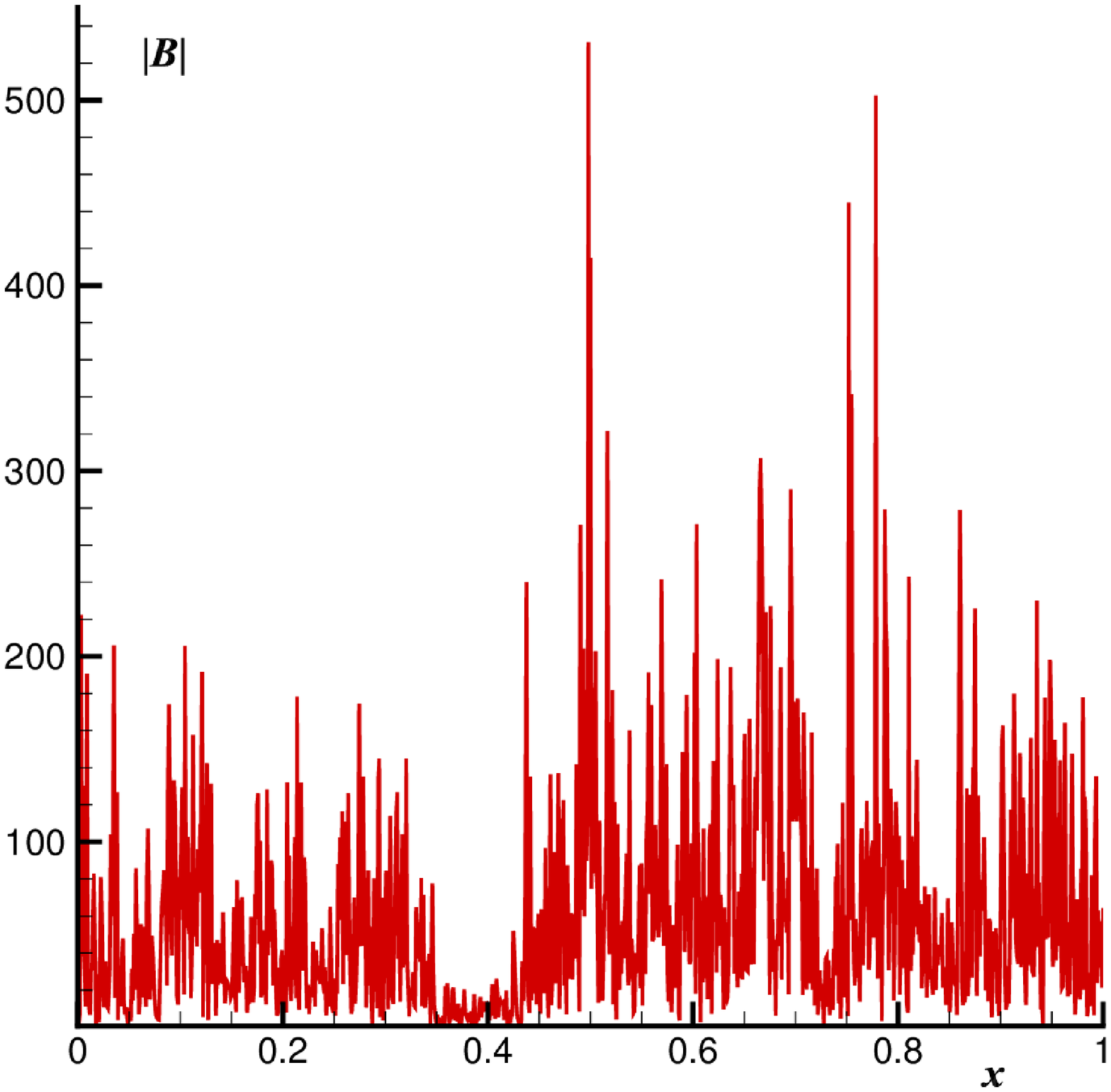}}
\caption{ Distribution of $|\mathbf{B}|$ along line $y=0.5$ at $t=100,220$.}
\end{figure}

These maxima define positions of the vorticity quasi-shocks which in the $k$-space, due to jets, are responsible for  anisotropy in
2D energy distribution $\epsilon (k_x,k_y)$.  Fig. 6 shows  
2D compensated  spectrum $k^4\epsilon (k_x,k_y)$ for the steady state
where one can see a set of jets with both strong and weak mutual overlapping.

\begin{figure}[t]
\label{Fig6}
\centerline{
\includegraphics[width=0.45\textwidth]{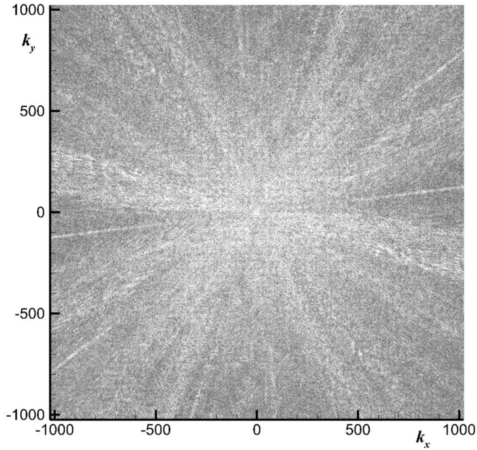}}
\caption{ 2D compensated spectrum $k^4\epsilon(k_{x},k_{y})$ at $t=220$.}
\end{figure}

Along each such jets $\epsilon\sim k^{-4}$ which after angular average leads with a high accuracy to the Kraichnan spectrum.

However, anisotropy, due to quasi-shocks, plays a more significant role for the higher moments of the velocity field than for the spectrum. Fig. 7 shows the dependencies of the third-order velocity structure function $S_3^{(L)}(\mathbf{R})=\langle
(\delta v_{\|})^3\rangle$, depending on $R$ for various values of angles, where $\mathbf{R}=\mathbf{r%
}-\mathbf{r}^{\prime }$ and $\delta v_{\Vert }$ the projection of the velocity
difference onto the vector $\mathbf{R}$.
At each angle $S_3^{(L)}(\mathbf{R})$ is close to the cubic parabola, i.e. $\propto R^3$, with a linear dependence relative $\eta$.
However, the average  of the third-order velocity structure function over angles only  gives a significant difference with the constant
$C_{3,isotr}=1/8$ for the isotropic turbulence. However, the angular averaging constant $C_3$ undergoes  temporal fluctuations: its  maximal value sometimes reaches 5  (for this <<steady>> state!). Average over time  within window $210\geq t \leq 465$ with characteristic period $\simeq 17$ gives better correspondence: $\bar C_3\simeq 2.4 C_{3,isotr}$.
\begin{figure}[t]
\label{fig7}
\centerline{
\includegraphics[width=0.5\textwidth]{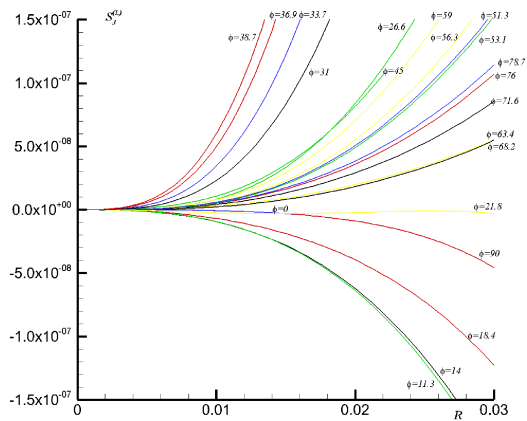}
}
\caption{ Dependence of $S_3^{(L)}$ as function of $R$ at different angles.}
\end{figure}

\section{Conclusion.}
The main resume of this work is the fact that  for the 2D turbulence, in the presence of pumping and damping, quasi-shocks,
originated because of compressibility  of di-vorticity lines,  play very essential role in the direct cascade. Quasi-shocks
are key factors which  define 2D turbulence anisotropy, in particular,  anisotropy  of both  spectrum and   structure functions.
Surprisingly that averaging over angles of 2D energy distribution $\epsilon (k_x,k_y)$ gives with a high accuracy the Kraichnan
spectrum. This result coincides with  results of the previous numerical experiments  \cite{Gotoh}-\cite{LindborgVallgren}. Note,
the same averaging for the third-order velocity structure function yields a value for constant $\bar{C_3}$ different from a pure
isotopic case  (see, e.g., review \cite{boffetta}
and references therein), in spite of the correct dependence of $S_3^{(L)}$ on both enstrophy flux $\eta$ and separation $R$.
This is an open question which we are going to study in our future work.

The authors thank A.N. Kudryavtsev and G.E. Falkovich for useful discussions. This work was
supported by the RSF (Grant No. 14-22-00174).

\end{document}